\documentclass[fleqn,10pt]{wlscirep}
\usepackage[utf8]{inputenc}
\usepackage[T1]{fontenc}
\title{Controllable shifting, steering, and expanding of light beam based on multi-layer liquid-crystal cells}

\usepackage{amsmath}
\usepackage{amsfonts}
\usepackage{mathrsfs}
\usepackage{siunitx}
\usepackage{bm}
\usepackage{hyperref}
\usepackage[]{units}
\usepackage{soul}

\graphicspath{ {./figures/} }

\author[1]{Urban Mur}
\author[1,2]{Miha Ravnik}
\author[1,3,*]{David Se\v{c}}

\affil[1]{Faculty of Mathematics and Physics, University of Ljubljana, Jadranska 19, 1000 Ljubljana, Slovenia}
\affil[2]{J. Stefan Institute, Jamova 39, Ljubljana, Slovenia}
\affil[3]{Adria Tehnik d.o.o., Na jasi 12, Tržič, Slovenia}

\affil[*]{david.sec@fmf.uni-lj.si}

\begin{abstract}
Shaping and steering of light beams is essential in many modern applications, ranging from optical tweezers, camera lenses, vision correction to 3D displays. However, current realisations require increasingly greater tunability and aim for lesser specificity for use in diverse applications.
Here, we demonstrate tunable light beam control based on multi-layer liquid-crystal cells and external electric field, capable of extended beam shifting, steering, and expanding, using a combination of theory and full numerical modelling, both for liquid crystal orientations and the transmitted light. Specifically, by exploiting three different function-specific and tunable birefringent nematic layers, we show an effective liquid-crystal beam control device, capable of precise control of outgoing light propagation, with possible application in projectors or automotive headlamps.
\end{abstract}
\begin{document}

\flushbottom
\maketitle

\thispagestyle{empty}

Diverse photonic applications rely on beam shaping and steering which is today at the macroscopic device level predominantly done with mechanically movable parts or prefabricated integrated circuits \cite{DavidsonN_ProgressinOptics45_2003,DuH_Micromachines7_2016,HeckMJ_Nanophotonics6_2017}. 
Alternative approaches also include using complex optical properties of different soft, solid, or composite materials, such as arrays of lenses and mirrors of specific shapes \cite{ChronisN_OptExpress11_2003,MatsuoS_ApplPhysA80_2005}, metasurfaces \cite{YuN_Science334_2011,GenevetP_Optica4_2017} or fibres \cite{LeeH_JLightwaveTechnol37_2019}.
Soft liquid crystalline (LC) materials are major optically active materials that exhibit tuneability due to their inherent optical anisotropy (i.e. birefringence) and ability to change their spatial birefringent profile with external fields, such as with electric field and confining surfaces~\cite{khoo2007liquid,Chigrinov_VG_LiquidCrystalPhotonicsNova}.

Liquid crystals are used in tunable lenses, where external electric field \cite{FrayA_InfraredPhysics18_1978,SatoS_JpnJApplPhys18_1979,TsengM_JApplPhys109_2011,FanF_OptLett38_2013,AlgorriJF_SciRep10_2020,JamaliA_OptExpress28_2020} 
or other external stimuli~\cite{BegelL_ApplOpt57_2018} are used to change optical properties of the lens.
LC lenses can also be generalised for use in beam shaping \cite{SioLD_ApplOpt57_2018} and beam steering \cite{MasudaS_ApplOpt36_1997,YinK_AdvOptMater8_2020}. 
Beam steering employs spatially varying director profiles to transform the refractive indices and thus guide light in a specific direction \cite{FrayA_ElectronLett11_1975,SasakiA_ElectronLett15_1979,MasudaS_ApplOpt36_1997,ApterB_OptEng44_2005,LaudynUA_SciRep7_2017,OtonE_JLightwaveTechnol37_2019}; however, with rather small steering angles ($< 10^{\circ}$) and rather high electric fields needed \cite{FrayA_ElectronLett11_1975,AlgorriJF_SciRep10_2020,TianL_OpticsCommunications481_2021}. Therefore, beam steering is often realised by \emph{reflection}, rather than \emph{refraction} from LC cells \cite{HeZ_Crystals9_2019}.
One of the used methods is constructing a grating that deflects light \cite{MaticRM_SPIEProceedings_1994,FanF_ApplPhysLett100_2012,OtonJM_JMolLiq267_2018}, where the deflecting angles can be up to $\unit[30]{^{\circ}}$ \cite{WengY_OptExpress24_2016,YinK_InformationDisplay37_2021} and efficiency can be even more than $90~\%$ \cite{GaoK_OptExpress25_2017,YinK_InformationDisplay37_2021}. Such devices are sensitive to incoming polarization and reflect light also in more than one diffraction order \cite{HeZ_Crystals9_2019}.
Another interesting example is to use LC elastomer fibres which rotate mirrors in order to deflect light into selected directions \cite{NocentiniS_SoftMatter13_2017}.
Overall, in most of the studied examples, the tuning of the beam steering and shaping is
rather limited (up to $\unit[8]{^{\circ}}$ continuously) \cite{MorrisR_JApplPhys126_2019} and devices are optimized for only two steering angles (diffractive orders) -- $0^{\circ}$ in off state and non-zero deflection in on state.

Achieving larger optical beam control by LC ordering can be done either by stacking LC cells in an array~\cite{MaticRM_SPIEProceedings_1994} or by utilising nonlinear optical effects~\cite{PecciantiM_Nature432_2004,AlberucciA_NewJPhys15_2013,AlberucciA_Optica2_2015,BeeckmanJ_OptExpress12_2004}.
For beam steering, multiple adjacent LC cells can be used to improve steering angle in a dual-twist Pancharatnam phase device \cite{ChengH_ApplOpt54_2015}, where deflection angles up to $80^{\circ}$ can be obtained and still maintaining very high efficiency.
Expanding these effects, wave front shaping and switching between different helical modes can be achieved by a cascade of Pancharatnam-Berry phase optical elements \cite{MarrucciL_ApplPhysLett88_2006}. Similarly, cascading multiple cycloidal
diffractive waveplates results in multiple diffraction angles \cite{ChenH_OptExpress24_2016}.
In nematic liquid crystals the interplay of material nonlinearity and optical intensity can lead to non-diffracting laser beam, called ``nematicons''. 
Higher power laser beams can realign the nematic director configuration and create a self-confining extraordinarily polarized laser beam with no diffraction, which propagates at a walk-off angle~\cite{PecciantiM_Nature432_2004}.
This angle can be tuned with reorientation of the bulk nematic configuration, either via electric~\cite{PecciantiM_NaturePhys2_2006} and magnetic fields~\cite{IzdebskayaY_NatCommun8_2017,PerumbilavilS_ApplPhysLett113_2018} or different colloidal inclusions~\cite{IzdebskayaYV_JOptSocAmB30_2013,IzdebskayaYV_OptLett39_2014}. Such steering of the beams was observed to achieve angles as large as $55^{\circ}$\,\cite{PiccardiA_ApplPhysLett100_2012,BarbozaR_OptLett36_2011}.

In this paper, we explore as the central scientific question the use of (three) close-stacked liquid crystal layers for light beam modulation and control, using combination of theory and full numerical simulations based on Landau-de Gennnes free energy minimisation, analytical Ansaetze and FDFD light simulations. Specifically, the three stacked LC layers each perform one designed function/modulation of the beam: one layer shifts the beam, second deflects the beam, and the third layer expands the beam, overall together, enabling an extensive beam modulation.
Such an array of building blocks can - by applying external electric field - change the incoming beam orientation by more than $25^\circ$ and focal spot continuously. The stacked liquid crystal device can also partition the incoming light beam into multiple beams (in multiple steps according to the number of building blocks). More generally, this work is aimed as a contribution towards possible experimental realisation of efficient, simple and wide-use control of light beams -- including steering, shifting, focusing and expanding -- which could be used in wide use applications such as projectors and automotive headlamps.

\section*{Results}

\begin{figure}[ht]
\centering
\includegraphics[width=0.35\linewidth]{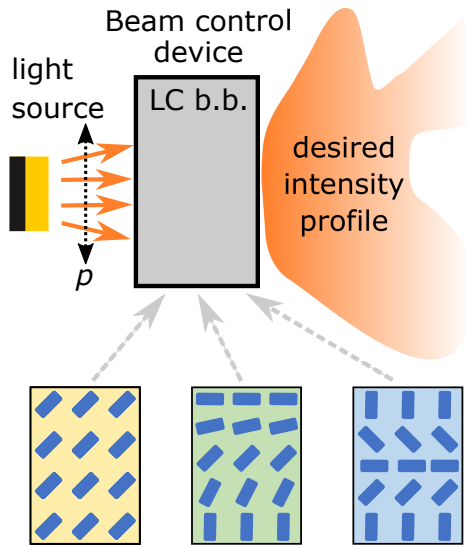}
\caption{Schematic view of a liquid crystal (LC) beam control device. The device is composed of a single or stacks of LC building blocks (LC b.b.) with some examples shown below. 
Blue cylinders represent nematic director and $p$ shows the polarization of the incoming light. 
The desired light profile is defined by locally tunable birefringent structure in the LC building blocks.}
\label{fig1}
\end{figure}

The beam control device (grey box in scheme in Fig.~\ref{fig1}) consists of multiple stacked building blocks (in principle, it can also be only one building block -- special liquid crystal cell), 
which can shape and transform the incoming light into a desired spatially varying intensity profile, by means of locally tunable LC birefringence.

The building block of such a device is a plan-parallel cell filled with a nematic liquid crystal.
To realise specific director field configurations, such as those shown in Fig.~\ref{fig1}, various external fields can be used like different anchoring types and strengths. 
In this work, nematic director profiles are both calculated by a full tensorial Landau-de Gennes free energy minimisation approach\cite{deGennesPG_1993, ravnik2009landau} and also
set by different analytical Ansatz functions mimicking either numerically or experimentally known fields. By using a free energy minimisation we show on a selected example how the used nematic director fields can possibly be created in practice by use of different anchoring types at the boundaries and applying static electric field as induced by electrodes with fixed electric potentials. Similar approaches have been used to explain and guide experiments in the past \cite{aguirre2016sensing, nikkhou2015light}. More details on the used methodological approach is given in Methods.
Experimentally, spatially varying 3D structures can also be stabilized by patterned anchoring, which can be achieved by use of various techniques, for example self-assembled monolayers \cite{gupta1997design, prompinit2010controlling}, photoalignment \cite{chigrinov2013photoaligning, nys2020patterned} or in-situ polymerisation \cite{tartan2018read}. 
Due to the inherent birefringence of the nematic medium, the local average molecular orientation -- the director field -- governs the optical properties of the material. For a polarization laying in plane defined by the optical axis, which is parallel to the nematic director, and the wave vector of light, the refractive index of the material depends on the angle between the wave vector and the optical axis -- nematic director field. Hence by rotating the nematic director field, optical properties of the material for a selected polarization (i.e. extraordinary beam) are tuned whereas remaining unaltered for the orthogonal polarization pointing out of plane (i.e. for ordinary beam).
The director configuration in each building block is taken as stationary, stabilized either by surface anchoring or appropriate external fields (e.g. electric).
The surface anchoring can be weak or strong, which in turn affects the needed strength of the external fields. Employing complex electric fields results in  non-uniform birefringent optical profiles. Here we present the beam control device that consists of three different building blocks to achieve extensive control over the light beam [as shown in Figure~\ref{fig2}(a)]:
(i) shifter, (ii) deflector and (iii) expander building block (lens).

\subsection*{LC beam control} 

\begin{figure}[!ht]
\centering
\includegraphics[width=0.65\linewidth]{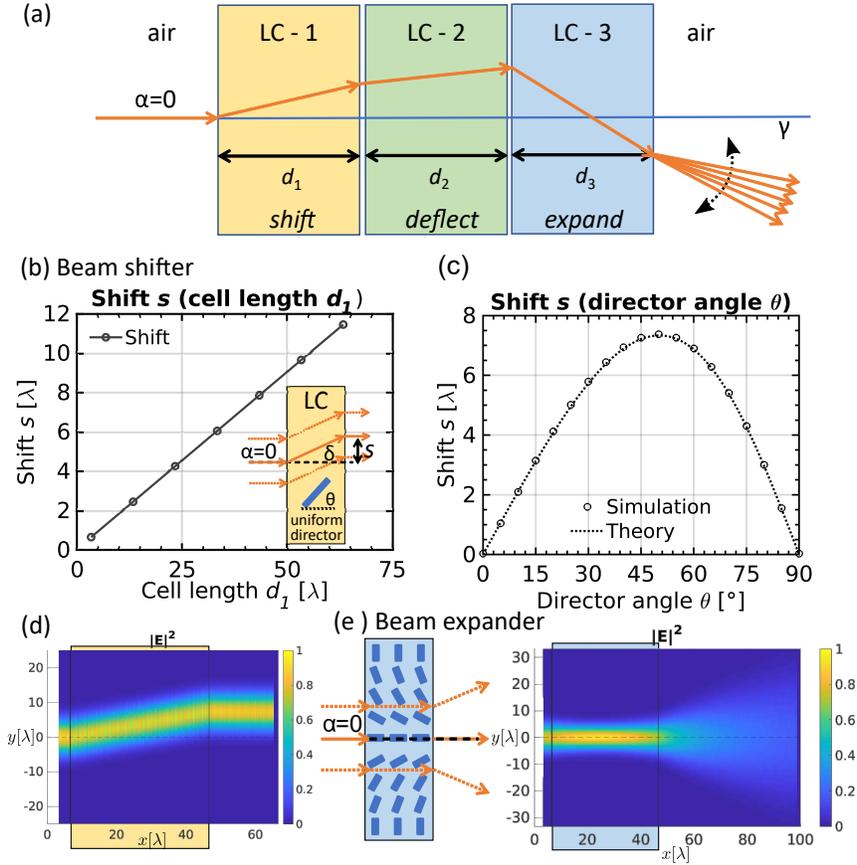}
\caption{Liquid crystal (LC) beam control: 
(a) A schematic representation of a stack of LC building blocks. Controlling the output beam for obtaining a desired beam profile and direction is done in three steps: (i) first the incoming beam is shifted, (ii)
then deflected to a certain angle and (iii) eventually expanded. Such an array of tunable building blocks can control
the output beam continuously and with great precision. 
(b) Dependence of the beam shift $s$ on the building block length $d_1$ for a fixed $\theta=45^{\circ}$. $\lambda$ is wavelength in vacuum. Beam shifter building block has a uniform LC director field (note the blue cylinder representing nematic director -- i.e. optical axis).
(c) Beam shift dependence on the LC director angle $\theta$. Maximum shift is
obtained, when the director is aligned at an angle of approximately $\unit[50]{^{\circ}}$. Length of the block was fixed at $d_1=40\lambda$. Theoretically predicted shift is obtained as $s=d_1\tan{\delta}$, where $\delta$ is a walk-off angle, given by Eq.~\eqref{walkoff}.}
(d) Simulated electric field intensity $|E|^2$ in a
shifter building block for an in-plane polarized Gaussian input beam. The beam is gradually shifted throughout the building block (note the yellow box representing
the shifter building block) and the shape of the beam is preserved. 
(e) An example of simulated electric field intensity $|E|^2$ for the expander building block: Incoming perpendicular Gaussian
beam is expanded in order to illuminate a larger area, using a radially escaped LC
director profile.
\label{fig2}
\end{figure}

A beam shifter (see scheme in Figure~\ref{fig2}b) is constructed of a LC cell with a uniform director field, which is oriented at an angle $\theta$ relative to the cell surface normal, uniformly across the whole cell of width $d_1$. 
In such a building block, the beam incident angle ($\alpha$) is the same as refracted angle as the phase front shape remains unchanged and the beam only gets shifted. 
The shift $s$, when director angle $\theta$ is constant, is dependent on the length of the building block $d_1$ and can be calculated as $s=d_1\tan{\delta}$, where $\delta$ is a walk-off angle obtained from Eq.~\eqref{walkoff} as presented in Figure~\ref{fig2}b for $\theta=45^{\circ}$.
The most relevant parameter for pre-positioning the beam for deflection is the beam shift $s$ at a fixed length of the block $d_1$. Its dependence on the director angle $\theta$ for $d_1=40\lambda$ is shown in Figure~\ref{fig2}c. For used material parameters, the maximal shift per unit length of the building block is at director angle $\theta\sim50^{\circ}$. Note that shifting also works for Gaussian beams with high waist-to-wavelength ratio (here at least 5:1) as shown in Fig.~\ref{fig2}d. The incident perpendicular Gaussian beam with the in-plane polarization gets gradually shifted along the building block length and the outgoing beam is again perpendicular to the building block, but shifted upwards by $s$ (see also inset in Fig.~\ref{fig2}b).
Such shifter building block is used to re-position or pre-position the beam relative to the profile in the deflector or expander building blocks, as shown later.

The third building block -- the expander -- is essentially a liquid crystal micro-lens, that expands the beam to a broader area (see Fig.~\ref{fig2}e).
We use a simple radially escaped $+1$ disclination line profile that is known to act as a diverging lens with positive birefringent nematics~\cite{LinH_MolCrystLiqCryst204_1991,CanculaM_OptExpress24_2016}, where the refractive index varies radially from the centre of the escaped line profile. The extraordinary beam acquires a phase shift which depends on the distance from the centre of the lens which can also be tuned by using external electric field~\cite{HarkaiS_PhysRevResearch2_2020}.
This allows us to continuously affect the focusing of light front as well as the effective focal length (numerical aperture) of the lens according to the desired application. We should comment that other LC cell lens profiles could be used such as hole-patterned microlenses, cylindrical and rectangular lenses, changing surface profiles of the cells, planar cells with floating-ring electrodes, etc.~\cite{AlgorriJF_Crystals9_2019,HonmaM_OptExpress17_2009,NoseT_JpnJApplPhys39_2000,HsuC_LiqCryst45_2017}.

\subsection*{Deflecting the beam}

\begin{figure}[ht]
\centering
\includegraphics[width=0.95\linewidth]{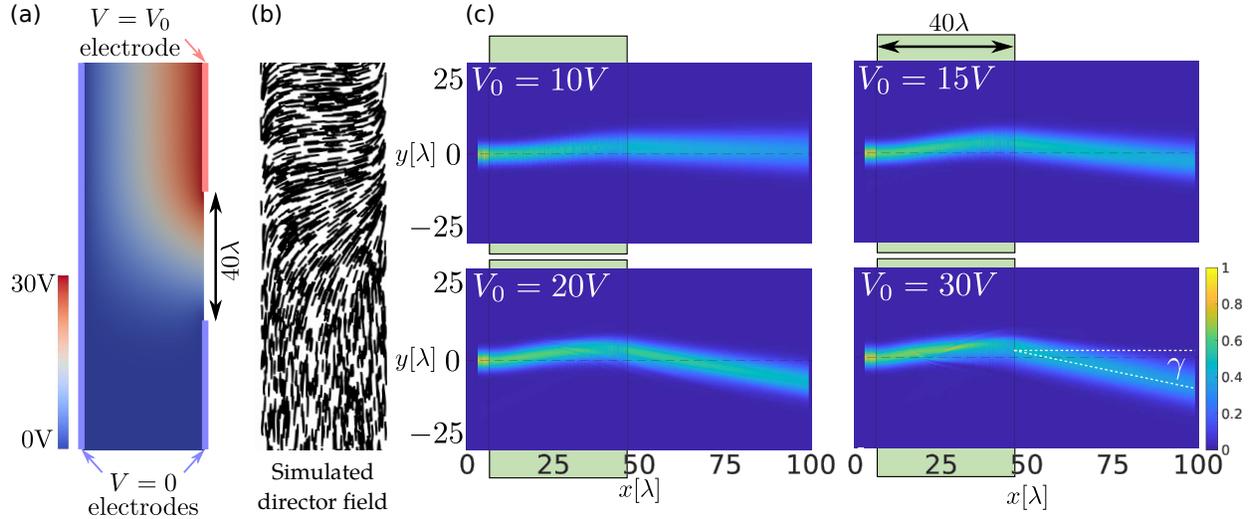}
\caption{Deflecting the beam with a planar LC cell and applied voltage. (a) Electrode distribution and simulated electric potential for $V_0=\unit[30]{V}$. (b) Simulated director field profile for electric potential shown in (a). Strong planar anchoring was used as a boundary condition on the electrodes. (c) Simulated light deflection by use of simulated director field for four different voltages at the electrode. Waist of the beam was set to $w_0=\unit[3.33]{\lambda}$, while the length of the building block was fixed to $d_2=\unit[40]{\lambda}$.}
\label{figv}
\end{figure}

The deflector building block [second building block in Fig.~\ref{fig2}(a)] can steer the incoming perpendicular beam continuously to a desired predefined angle $\gamma$ as shown in Fig.~\ref{figv}(c).
We modelled such cell profile using Q-tensor Landau-de Gennes free energy minimisation approach~\cite{ravnik2009landau} to numerically calculate the ordering of a nematic liquid crystal in the presence of electric field (electric potential), induced by electrodes, as presented in Fig.~\ref{figv}(a,b). We assumed three electrodes -- one on the incoming side of the LC cell and two on outgoing side [see in Fig.~\ref{figv}(a)]. By applying different voltages on one electrode, various clinotropic (bent-aligned) director field configurations are obtained by means of free energy minimisation. Note that by changing the voltage $V_0$ on the electrode [Fig.~\ref{figv}(c)] deflection angle can be controlled.

\begin{figure}[ht]
\centering
\includegraphics[width=0.95\linewidth]{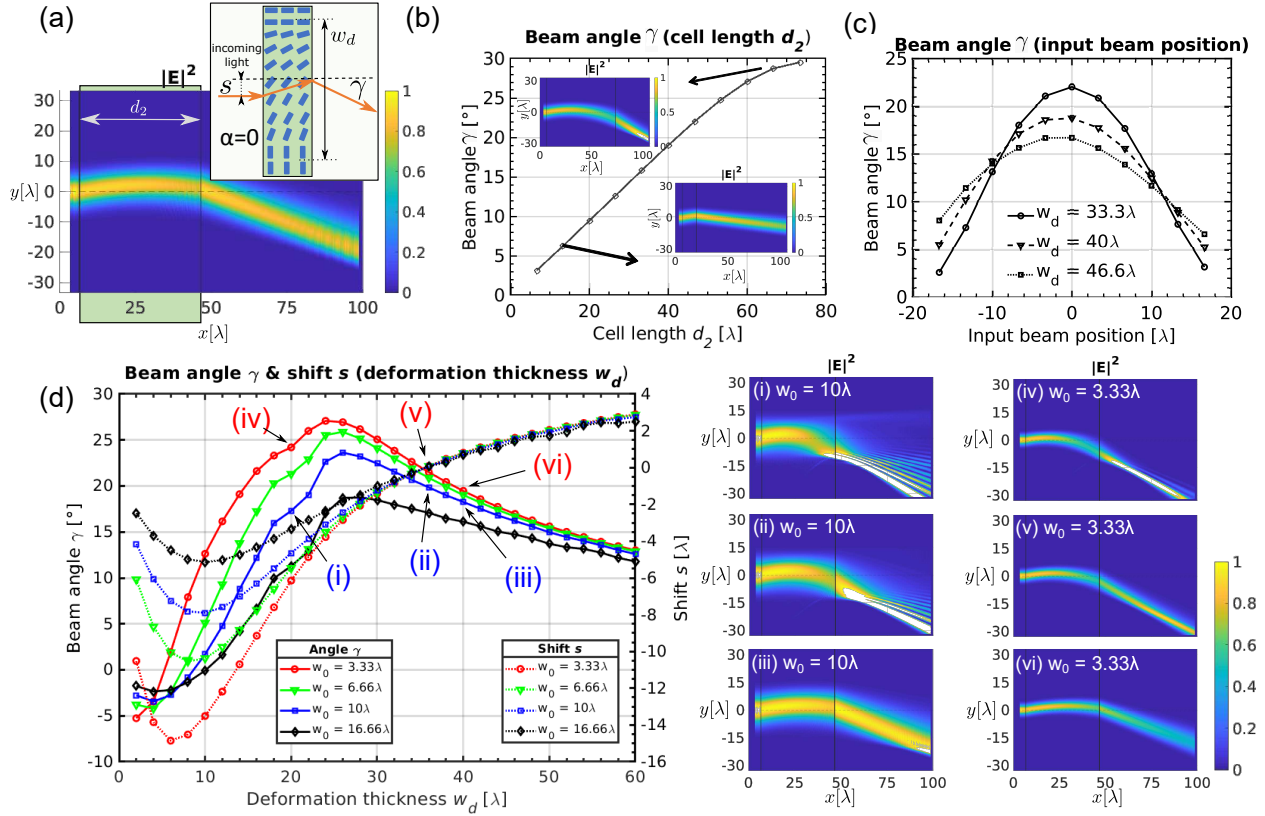}
\caption{Deflecting the beam. 
(a) The incoming Gaussian beam, which is normal to the building block, is deflected at an
angle $\gamma\sim\unit[20]{^{\circ}}$ and shifted upwards.
Inset shows schematic representation of the building block with the deformation thickness $w_d$ which shifts the incoming beams for the distance $s$ and deflects it to an angle of $\gamma$ as shown with simulated normalized electric field intensity $|E|^2$. 
The beam deflection angle $\gamma$ can
be tuned by changing (b) the building block length $d_2$, (c) position of the input beam or the (d) deformation thickness $w_d$.
(c) Dependence of the beam angle $\gamma$ on the position of the incoming beam.
Both (b) and (c) are calculated for $w_0=\unit[6.66]{\lambda}$ and $w_d=\unit[40]{\lambda}~/~d_2=\unit[40]{\lambda}$, respectively.
(d) Changing the deformation thickness $w_d$ with respect to beam waist $w_0$ greatly affects the shift $s$ and deflection angle $\gamma$
and can be tuned with external electric fields. Length of the building block was fixed to $d_2=\unit[40]{\lambda}$.}
\label{fig3}
\end{figure}

To study different material parameters and properties of such cells, we use a LC cell with clinotropic (bent-aligned) director field inside the building block as induced by electric field. 
In this deflector geometry, the rate of change $K$ of the director angle in the lateral direction (i.e. perpendicular to the incoming wave vector) and the building block length -- together with the material parameters of refractive index birefringence, elastic constant and surface anchoring strength -- control the deflection angle.
In a longer building block [see Fig.~\ref{fig3}(b)],
the deflection angle is larger as bigger phase shift accumulation difference between upper and lower end of the beam leads to wave fronts inclined at a larger angle.
The relation is rather linear and by changing the building block length angles up to $\sim\unit[30]{^{\circ}}$ are achievable. 

Alternatively to changing the actual building block length, the beam angle can be tuned by shifting the position of the input beam up or down [Fig.~\ref{fig3}(c)] or by changing the rate of change of the director angle $K$ [Fig.~\ref{fig3}(d)], which is here characterized by the deformation thickness $w_d \propto 1/K$ -- the lateral distance within which the director turns for $90^{\circ}$, from parallel to perpendicular orientation with respect to the building block surface normal [see Fig.~\ref{fig3}(a)].
Note in Fig.~\ref{fig3}(c), that for a small variation in initial position of the beam from the centre (position = 0 is in the middle of the bend region), the deflection angle changes only slightly.
Mainly, by changing the deformation thickness $w_d$ three different regimes can be observed with respect to its ratio to beam waist thickness $w_0$ ($w_0/w_d$).

When beam diameter $2w_0$ is comparable to the deformation thickness $w_d$ [$w_d\sim 2w_0$, see Fig.~\ref{fig3}(d), insets (i) and (ii)], we observe large deflection angles but also beam splitting. In this regime, the bend area is narrow and there are steep changes in the refractive index profile which leads to different parts of the incoming beam following different diffraction paths, in turn splitting the beam into multiple high- and low- intensity regions. In actual experimental setting, such beam splitting will be strongly affected (and will likely disappear) due to light scattering and defocusing caused by thermal fluctuations of the nematic director and illumination with a non-coherent light will blur the output signal~\cite{HessAJ_PhysRevX10_2020}. Note that the beam gets also shifted upwards when travelling through the building block.

As we gradually increase the deformation thickness towards the values greater than the beam diameter [$w_d > 2w_0$, see Fig.~\ref{fig3}(d), inset (iii) and inset (iv) for a beam with smaller waist] the largest deflection angles occur. However, despite eliminating the splitting, brighter areas appear after the deflection and the beam is still not entirely uniform.

With a larger deformation thickness ($w_d \gg 2w_0$) we encounter a linear regime, where deflection angle is roughly linearly dependent on the deformation thickness ($\gamma \propto w_d$) [see insets (v) and (vi) in Fig.~\ref{fig3}(d), where  $w_d\geq 10w_0$], and notably deflection angles of up to $\gamma= 20^\circ$ can still be achieved. Additionally, with $w_d \gg w_0$, there is only weak dependence on the waist thickness $w_0$, which opens further application possibilities as beams of different sizes and shapes can be mutually controlled. Note also that in this regime (iii) the shift $s$ is close to zero which can be particularly useful as there is no need for an additional shifter building block to eliminate the shift.

\begin{figure}[ht]
\centering
\includegraphics[width=0.42\linewidth]{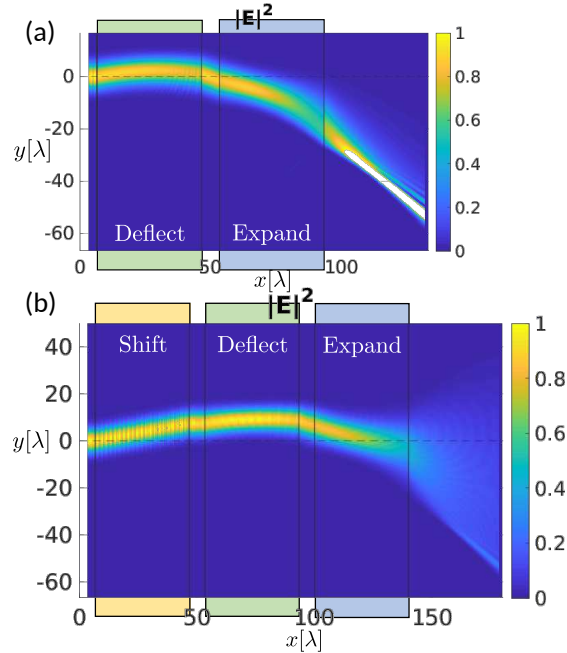}
\caption{Stacking of building blocks into a multi-layer beam control device. (a) Simulated electric field intensity $|E|^2$ for
a double building block device consisting of a deflector and an expander. 
The beam is efficiently deflected, but expanded non-uniformly, since it does not pass through the centre of the expander block (dashed line). (b) Simulated electric field intensity $|E|^2$ for a triple building block device consisting of a shifter, a deflector and an expander. The shifter pre-positions the beam and thus ensures that it passes through the centre of the expander (dashed line). Uniformity of the expansion is improved. All building blocks in both panels are \unit[40]{$\lambda$} long with the in-between spacing of \unit[6.67]{$\lambda$}.}
\label{fig4}
\end{figure}

Combining several LC building blocks -- i.e. forming stacks of LC cells --  results in a tunable beam control device, capable of different manipulations of the beam. As already presented in Fig.~\ref{fig3}(c), tuning of the deflection angle can be achieved by a combination of a shifter and a deflector: by varying the director angle in the shifter building block, the position of the incoming beam on the deflector can be altered and as a result, the beam deflects to a different angle.
Experimentally, tuning the director field configuration can be achieved by locally modulating electric field in the liquid crystal cell, for example by using electrodes on the surface of the cell as shown in Fig.~\ref{figv}(c).
Similar setup, presented in Figure~\ref{fig4}(a), can be used to expand the deflected beam. In such setup, the beam can be controlled in two ways: (i) tuning the deflection angle via deflector parameters as presented in the previous section and (ii) tuning the illuminated area by changing the lens power. 
However, by expanding the deflected beam, some brighter areas appear. Since the deflected beam does not travel through the centre of the expander, but more through the bottom part, the block, as it has an escaped disclination profile, then acts additionally as a deflector and expands the beam non-uniformly, resulting in much brighter spots at the bottom part.

This non-uniformity of the outgoing light can be addressed by adding another building block -- a shifter -- that compensates for the shift due to propagation at an angle after deflection [see Figure~\ref{fig4}(b)].
The incoming beam is thus firstly lifted upwards in order to pass the lens through its centre. Note, that this creates a more uniform output beam even if the lens is unchanged, but reduces the control of the deflection angle of expanded beam since shifting the beam also slightly changes the deflection angle as presented in Figure~\ref{fig3}(c). 
The shifter building block could be incorporated in the deflector cell, when precise local director control is possible. Understanding its own effect can also help in designing such a device.
Overall, the presented mutual tuning of nematic birefringence fields in the different building blocks results in a simple and precise tunable micro device capable of controlling the incoming light direction and shape.

\subsection*{Colour tuning - spectral dependence}

\begin{figure}[htb]
\centering
\includegraphics[width=0.95\linewidth]{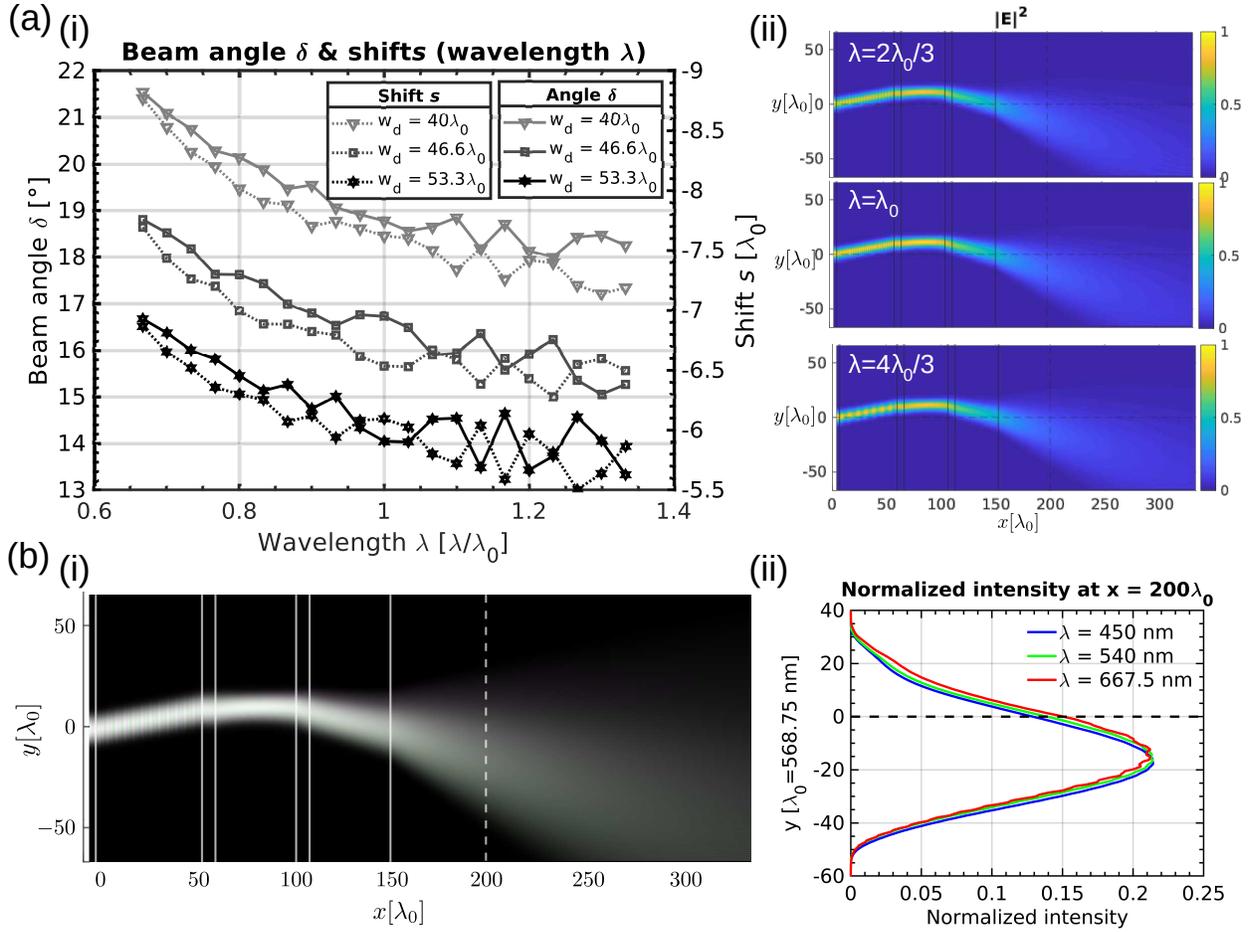}
\caption{Colour  tuning in 3 layer stacked device.
(a) Wavelength dependence of the beam control with a three-building block device. (i) There is a slight change of shift and deflection angle produced by the deflector building block when the wavelength of the light is changed. (ii) By optimizing the device parameters for a central wavelength  of the input spectrum, the effects can be minimized and could be further improved by selecting a LC with appropriate 
dispersion properties.  
(b) The propagation of white light emitted by an RGB light source into the beam control device. Note good beam colour control. Logarithmic values of RGB intensities are used to plot panel (i). There is good alignment of the intensity peaks of three different wavelengths further away from the lens (at the position, marked by dashed line) as shown in (ii). 
}
\label{fig5}
\end{figure}

Different beam control applications rely on non-monochromatic light, so we additionally explored the effect of different wavelengths of incoming beam -spectral dependence- in the performance of the beam control device. We varied the wavelength of the incoming beam by $\unit[\pm30]{\%}$ to cover a broader light spectrum - this can roughly cover the whole visible light (i.e. from $\unit[380]{nm}$ to $\unit[750]{nm}$).
We particularly focused on the deflector [Figure~\ref{fig5}(a)] and analysed the deflected beam angle $\gamma$ and corresponding shifts $s$ with respect to the incoming beam wavelength.
By changing the wavelength, both deflection angle and shift change, but the change is roughly linear and only around $\sim\unit[4]{^{\circ}}$ for a $\sim\unit[60]{\%}$ change in wavelength. 

To emulate the white light (i.e. broader wavelength light) passing through beam control device, we generate an RGB superposition of three beams with different wavelengths: (i) $\lambda_1=\unit[450]{nm}$, (ii) $\lambda_2=\unit[540]{nm}$ and (iii) $\lambda_3=\unit[667.5]{nm}$. 
This beam then passes through a device with a maximal deflection angle, already shown in Figs.~\ref{fig4}(b),\ref{fig5}(a). We observe good colour robustness of the beam manipulation device [see Fig.~\ref{fig5}(b)], with only minor colour-dispersion and an expanded cone of light that is propagating at an angle with respect to the input beam. Overall, the results show that the presented beam manipulation device can be used for a manipulation of broad wavelength light beams.

\subsection*{Tuning of the beam control device}

By tuning the liquid crystal orientation in the shifter and the deflector different - continuous and tunable - beam deflection angles and beam expansion can be achieved. 
The desired beam angle is selected by realising the proper deformation thickness $w_d$ and its orientation in the deflector (for deflecting either up or down). The maximum deflection angle is pre-determined by lengths of shifter and deflector building blocks.

\begin{figure}[!htb]
\centering
\includegraphics[width=0.7\linewidth]{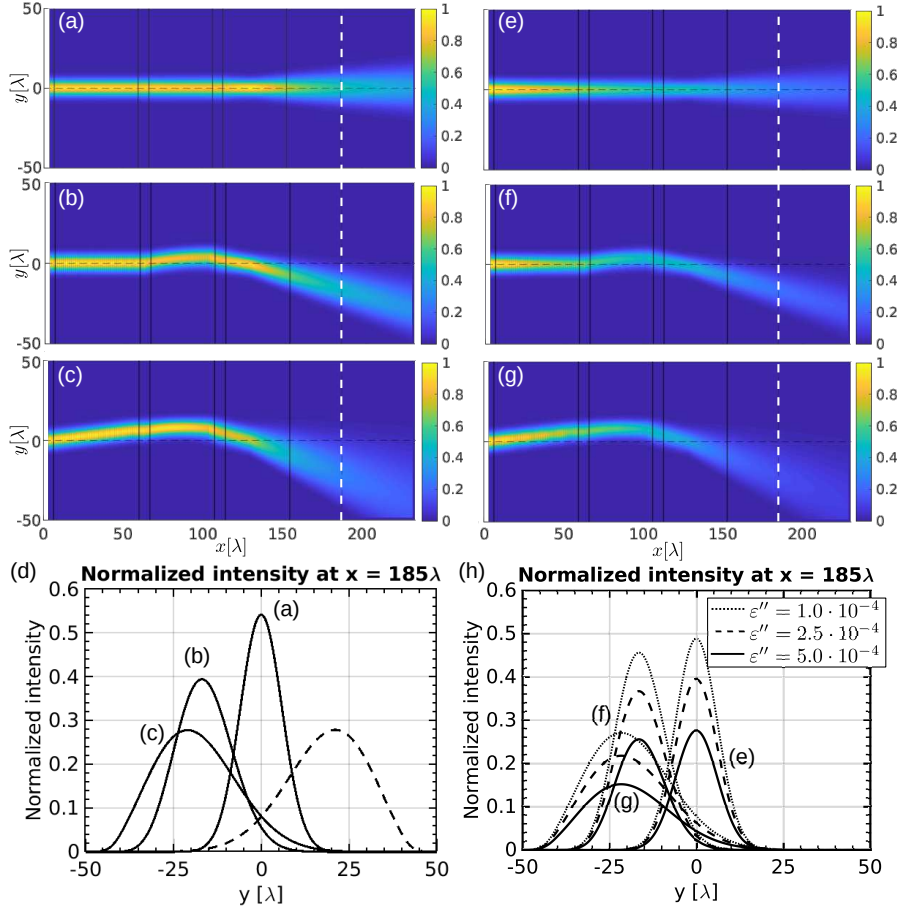}
\caption{Tuning of beam control device. (a) LC beam controller in an ``off'' state. The LC director field is
homogeneous in first (shift) and second (deflector) building blocks, so there is no shift or deflection. The total deflection angle is thus $\gamma = \unit[0]{^{\circ}}$ and the beam is only expanded. 
(b) Medium beam deflection of $\gamma\sim\unit[15]{^{\circ}}$. The deflector deformation width is set to $w_d = 66.6\lambda$ to achieve the desired deflection angle 
for a selected beam waist (see Fig.~\ref{fig3}) and the shifter building block angle is kept at $\theta = \unit[0]{^{\circ}}$ because the beam will already get shifted enough by deflector.
(c) Large beam deflection. The deflector deformation width is narrowed to $w_d = 40\lambda$ to achieve larger deflection angle and the shifter building block is adjusted to the angle  $\theta\sim\unit[23]{^{\circ}}$ to compensate for a larger decentering of the beam due to larger deflection angle. The beam is expanded fully downwards to an angle of $\gamma\sim\unit[20]{^{\circ}}$. (d) Intensity profiles of all three beams at the position marked with the vertical dashed line in panels (a-c). Dashed line represents the profile of the beam after large deflection in the opposite direction. Expander building block (lens) remains unchanged for all cases. (e-g) Propagation of the beam through the lossy material with the same parameters as in (a-c), respectively. Complex part of permittivity was set to  $\varepsilon''=5\cdot10^{-4}$. (h) Intensity profiles of all three beams at the position marked with the vertical dashed line in panels (e-g) for different values of $\varepsilon''$.}
\label{fig6}
\end{figure}

Then the director angle in the shifter is selected so that the beam passes through the centre of the expander building block and is uniformly expanded. Notably, the deflection angle dependence on the position of the beam centre (i.e. shift produced by the shifter building block) presented in Fig.~\ref{fig3}(c) needs to be taken into the account. 
Subsequent angle changes due to shifting the beam can be minimized by using thinner expander building block, so that smaller shift is needed to ensure a more uniform expansion. 
The expansion itself could be further improved by optimizing and tuning the lens. Figure~\ref{fig6} presents three cases of beam control: Fig.~\ref{fig6}(a) shows non-deflected beam that is only expanded [see the peak at $y=0$ in Fig.~\ref{fig6}(d)], and Figs.~\ref{fig6}(b),(c) show beam deflection for approximately $\unit[15]{^\circ}$ and $\unit[20]{^\circ}$, respectively.
Additionally, we show the intensity profile [Fig.~\ref{fig6}(d), dashed line] for a beam deflected upwards. In Figs.~\ref{fig6}(e)-(h) we show that the absorption -- losses -- of the device do not change its characteristics, rather than just the magnitude of the transmitted intensity. Propagation with absorption was calculated by adding of isotropic complex permittivity $\varepsilon''$ of different magnitudes to the liquid crystal dielectric tensor. Significant deviations ($>\unit[10]{\%}$) in the magnitudes of output intensity profiles only occur when the values of $\varepsilon''$ are in the order of $10^{-4}$ to $10^{-5}$, while typically values $\varepsilon''$ in the liquid crystals are in the order of $10^{-7}$\,\cite{wu1998absorption}.

\subsection*{Partitioning the beam}

\begin{figure}[htb]
\centering
\includegraphics[width=0.9\linewidth]{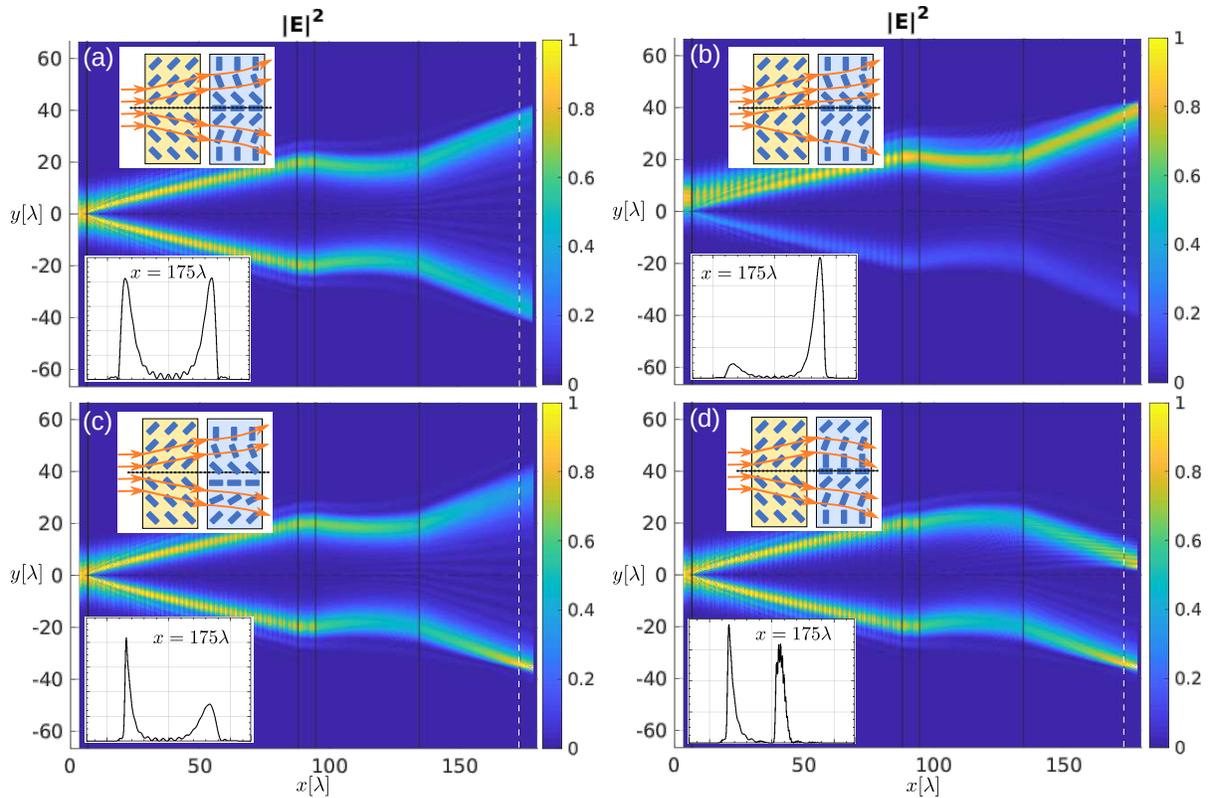}
\caption{Splitting and controlling the beam. (a) Two shifters can be used as a beam splitter: the beam is passed along the border between two areas with different (uniform) nematic director orientation. Separated beams can be deflected at the same or different angle, each with its own deflector or by lens. (b) By tuning the position of the splitting area of the director field, intensity in each part of the beam can be determined.  (c) If the centre of the lens does not coincide with the plane of splitting, the separated beams hit the lens at different distances from its centre and are deflected at different angles. (d) Split beams can be deflected in the same direction by using a pair of deflectors. }
\label{fig7}
\end{figure}

More complex intensity profiles can be obtained by splitting the incoming beam and controlling each part of the beam separately. In Figure~\ref{fig7}, we show beam splitting into two beams by using a double-shifter building block [see yellow inset in Fig.~\ref{fig7}(a)] with the director angle of the opposite sign in the upper and lower portion of the block. Deflecting each split beam is done by two deflectors, one on top of the other. An expander block, which essentially acts as a pair of deflectors with the opposite orientation, can be used if the deflection in the opposite direction is desired [see inset in Fig.~\ref{fig7}(a)]. By tuning the position of the splitting area of the director field, which equals moving the double-shifter from Figure~\ref{fig7}(a) up or down, intensity in each part of the beam can be determined [Fig.~\ref{fig7}(b)]. Additionally, the beams can be deflected at different angles by shifting the expander block up or down, relative to the centre of the beam [Fig.~\ref{fig7}(c)] or can propagate along the same direction, being parallel to each other, if an actual double deflector block is used [Fig.~\ref{fig7}(d)]. Moreover, each beam could further be split or expanded by using additional building blocks.


\section*{Towards experimental realisation}

The specific experimental approach we see as exciting for realisation of the proposed multi-layer liquid-crystal cells is to use the Two-Photon Polymerisation Direct Laser Writing (2PP-DLW), an emerging processing technique, which can fabricate polymer structures on the micro and even nanoscales \cite{o2020electrically, gan2013three, tartan2018read}. Such procedure could allow not only for fabrication of thin walls of polymer-networked liquid crystals with the thickness of as small as $\unit[1]{\mu m}$ \cite{o2020electrically}  between different elements of the cell (i.e. building blocks of our device  -- the shifter, deflector and expander), but could also realise highly-diverse spatially varying effective surface anchoring profiles on these printed separating walls, overall allowing  precise and customized prefabrication of different devices. Because of their small thickness, notably, the polymer walls would enable smooth transition of light between building blocks without too much undesired diffraction or shift, that would otherwise occur in a usually much thicker ($\approx\unit[100]{\mu m}$) glass separators. We note that in our work we used exemplary values of the material and geometric parameters; therefore, any optimisation could further  improve the performance of our device.

More generally, different experimental realisations of the beam control elements were reported that relate to our work. Beam shifting based on walk-off was realised experimentally in static \cite{piccardi2010power}, voltage \cite{PecciantiM_Nature432_2004} or magnetic field controlled \cite{IzdebskayaY_NatCommun8_2017, PerumbilavilS_ApplPhysLett113_2018} planar liquid crystal cells. Similarly, beam steering was reported with voltage driven  device \cite{BarbozaR_OptLett36_2011}, where self-guiding and solitonic states can be achieved with higher power. Beam deflection by voltage driven bent-align cell was achieved in positive \cite{FrayA_ElectronLett11_1975} and negative \cite{SasakiA_ElectronLett15_1979} birefringence liquid crystal cells. Different experimentally realisable structures for electrically tunable LC lenses were reported \cite{lin2011review, lin2017liquid}, such as curved lens \cite{kaur2016graphene}, gradient index (GRIN) lens \cite{ye2002optical}, Fresnel type lens \cite{huang2017polymer}, multi-layered lens \cite{chen2015polarizer} and polymer dispersed liquid crystal PDLC lens \cite{lin2011polarization}. Overall, there is exciting -experimental and theoretical- progress realising beam and light steering with liquid crystals, typically relying on manipulation of single liquid crystals layers, which could be adapted and used for design of also  multi-layered nematic devices.

\section*{Discussion}

In this work we show tunable beam control device capable of controlling the outgoing light intensity propagation direction and profile with great precision.
The beam control device is based on multiple stacked nematic liquid crystal cells -- building blocks. 
To emphasize and demonstrate the fundamental beam control, we used rather simple building blocks -- shifter, deflector, and lens cells, but there is no principal limitation to use more advanced elements with multiple beam control functions or more elements .  
In the demonstrated approach the mutual tuning of each building block contributes to the total effect on the incoming beam. The beam can be deflected continuously in arbitrary direction (up or down in our case) and is then expanded (or focused) to provide a desired intensity profile. 
The beam can also be split into multiple sub-beams and each sub-beam can be controlled individually -- its direction, intensity and profile. 
To obtain a full 3D-dimensional beam control, which is beyond the scope of this paper, additional deflector building blocks could be implemented, in combination with a polarization rotator (for example half-wave plate), to steer the beam in different directions of the solid angle. 

The presented beam control device is -- with properly designed material parameters -- capable of controlling a broad-wavelength light with only little colour dispersion, which opens additional possibilities for applications.
Furthermore, by adding additional building blocks, more complex intensity profiles can be obtained. 
Future research will be directed to optimize the material parameters and find the optimal building block structures to 
simplify the realisation of such an adaptive beam control device.

\section*{Methods}
\subsection*{Light simulations}

Nematic materials are  used in optical applications importantly due to their ability to control the light via direction dependant refractive index --- i.e. the birefringence, which originates from the orientational organization of molecules along a preferred direction, called director (equivalent to the optical axis)~\cite{deGennesPG_1993}. The local orientation of the director can be widely tuned  with external fields, such as confining surfaces, electric or magnetic fields~\cite{deGennesPG_1993}. Poynting vector of a light beam is generally not parallel to the wave vector when it travels through uniaxial birefringent material~\cite{peccianti2005observation}. The walk-off angle $\delta$ between the wave vector and the Poynting vector for a polarization laying in the plane of the optical axis can be expressed via index ellipsoid \cite{bregar2017refraction} as:
\begin{equation}
    \tan{\delta}=\dfrac{(1-\frac{n_o^2}{n_e^2})\tan{\theta}}{1+\frac{n_o^2}{n_e^2}\tan^2{\theta}}
    \label{walkoff}
\end{equation}
where $\theta$ is the angle between the wave vector and the optical axis and $n_o$ and $n_e$ are the ordinary and extraordinary refractive indices of the birefringent medium. 
In addition to beam intensity modulation and relocation, also the phase profile of the beam can be altered via birefringence, for example by changing the angle between the wave vector and the optical axis. Therefore both the phase and intensity profile of the input beam can be controlled by LC nematic director configuration. 

The full vectorial control over the shaping of the light beams with the liquid-crystal beam control device is explored by using Finite-Difference Frequency-Domain (FDFD) numerical modelling based on solving the matrix form of the Maxwell curl equations  written in the frequency
domain:
\begin{equation}
((\nabla \times \bm{\varepsilon}^{-1} \nabla \times) -\omega^2 \bm{\mu} )(\vec{H}) = \vec{S}
\label{maxeq}
\end{equation}
where $\vec{H}$ is nodal magnetic field vector, $\vec{S}$ is a nodal source vector, $\omega$ is frequency of the light and $\bm{\varepsilon}$ and $\bm{\mu}$ are space-dependant matrices of material parameters in the units of $\varepsilon_0$ and $\mu_0$, respectively. Dielectric permittivity tensor $\bm{\varepsilon}$ depends on the local orientation of the optical axis, which is parallel to the nematic director. Director field profile is included in the nematic order parameter tensor $\bm{Q}_{\mathrm{ord}}$:
\begin{equation}
    Q_{\mathrm{ord}ij} = \frac{S}{2}\left(3n_in_j-\delta_{ij}\right)+\frac{P}{2}\left(e_i^{(1)}e_j^{(1)}-e_i^{(2)}e_j^{(2)}\right)
    \label{orderQ}
\end{equation}
where $S$ is the degree of order and $n_i$ are components of the nematic director which was either determined analytically or by use of minimisation of Landau-de Gennes free energy as explained below. $\vec{e}^{(1)}\perp\vec{n}$ is the secondary director and $\vec{e}^{(2)}=\vec{n\times}\vec{e}^{(1)}$. The second term in Eq.~\eqref{orderQ} accounts for biaxiality $P$, which quantifies fluctuations around the secondary director $\vec{e}^{(1)}$. $\bm{\varepsilon}$ is calculated from $\bm{Q}_{\mathrm{ord}}$ as \cite{deGennesPG_1993}:
\begin{equation}
\bm{\varepsilon}=\Bar{\varepsilon}\bm{I}+\frac{2}{3}\varepsilon_{\mathrm{a}}^{\mathrm{mol}}\bm{Q}_{\mathrm{ord}},
\label{epsq}
\end{equation}
where $\Bar{\varepsilon}$ is the average dielectric permittivity and $\varepsilon_{\mathrm{a}}^{\mathrm{mol}} = (\varepsilon_{\parallel}-\varepsilon_{\perp})/S$ is the molecular dielectric anisotropy for a degree of order $S$ and are related to the refractive indices of birefringent material at a given temperature, which were extracted from the literature \cite{li2004temperature}.
Grid spacing of at least $\lambda/10$ is used, where $\lambda$ is the wavelength of light in vacuum. The source vector is calculated using the total-field/scattered-field (TF/SF) formulation \cite{rumpf2012simple} as:
\begin{equation}
\vec{S} = (\bm{Q}\bm{A}-\bm{A}\bm{Q})\vec{\Tilde{S}}
\end{equation}
where $\bm{A}=((\nabla \times \bm{\varepsilon}^{-1} \nabla \times ) - \omega^2 \mu)$ is the wave matrix from Eq.~\ref{maxeq}, $\bm{Q}$ is the masking matrix, denoting the areas where total or scattered field is to be calculated and $\vec{{\Tilde{S}}}$ is the source field, propagating through vacuum. Perfectly matched layer (PML) with the thickness larger than $\lambda/2$ is used to truncate the domain and simulate infinite boundary conditions in all directions.

The solution of such linear system is the full vector field $\bm{H}$ consisting of a total magnetic field in the regions where the elements of masking matrix $\mathbf{Q}$ equal to zero and scattered magnetic field in the regions where they equal to one. Following the Maxwell equations, electric vector field in every point is obtained as
\begin{equation}
\vec{E} = \dfrac{1}{\omega}\varepsilon^{-1}\nabla \times \vec{H}
\end{equation}

\subsection*{Liquid crystal free energy minimisation and electric potential calculations} 

Minimisation of Landau-de Gennes free energy \cite{ravnik2009landau} was used to numerically calculate ordering of a nematic liquid crystal in the presence of electric field, induced by electrodes, as presented in Fig.~\ref{figv}. In addition to the Landau expansion, describing the temperature-driven phase transition, the free energy expression consisted of a single elastic constant approximation ($K_{\mathrm{el}}=$\unit{$1.264\cdot10^{-11}$}{N}) elastic free energy used to describe nematic distortions and the term describing the coupling with the static electric field. Static electric field was obtained as a gradient of electric potential $V$ which was determined by numerically solving the analogue to the Laplace equation in an anisotropic dielectric medium
\begin{equation}
    \partial_i ( \epsilon_{ij} (\partial_j V) ) = 0,
\end{equation}
where $\epsilon_{ij}$ are the components of dielectric tensor, by applying boundary conditions set by voltage on the electrodes. The minimum of the free energy was found via solving the Euler-Lagrange equations, while simultaneously relaxing the electric potential. A finite difference based explicit relaxation method was used. During the relaxation the Q-tensor and electric potential $V$ on all lattice sites were updated in each time step until the steady state was achieved, normally after $1\cdot10^5$--$2\cdot10^5$ relaxation steps.

\subsection*{Ansaetze for liquid crystal birefringence profiles} 

Birefringence profiles (i.e. the director field) in the liquid crystal (LC) cells for calculations presented in other figures were determined by analytical formulas 
\begin{align}
    \vec{n}_{\mathrm{shift}} &= (\cos\theta,\sin\theta,0), \qquad \theta = \mathrm{cons.} \\
    \vec{n}_{\mathrm{def.}} &= (\cos\theta(y),\sin\theta(y),0), \qquad \theta(y) =
    \begin{cases}
        0 & y \geq w_d/2 \\
        \frac{\pi}{2} \left(\frac{1}{2}-\frac{y}{w_d}\right) & w_d/2 > y > -w_d/2 \\
        \pi/2 & y \leq -w_d/2 
    \end{cases}\\
    \vec{n}_{\mathrm{exp.}} &= (\cos\theta(y),\sin\theta(y),0), \qquad \theta(y) =
    \begin{cases}
        \pi/2 & y \geq w_d/2 \\
        \pi(1-\frac{y}{w_d}) & w_d/2 > y > 0 \\
        -\pi\frac{y}{w_d} & 0 \geq y > -w_d/2\\
        \pi/2 & y \leq -w_d/2 
    \end{cases}
\end{align}
for a shifter, deflector and expander, respectively, where $w_d$ is the deformation thickness in the block. Also these director field configurations could be obtained from full numerical simulations  with free energy minimisation.

Computational domain was restricted to two dimensions due to high computer memory (RAM) consumption. The derivatives in the third dimension were eliminated, meaning that the obtained results are invariant in that particular direction. Modulation of the beam is therefore done in 2D only, but could in principle be extended to 3D by use of more complex 3D nematic director field profiles in individual cells or stacking multiple cells with orthogonal orientations. 2D configuration allowed us to simulate cells with the sizes of tens of wavelengths, i.e. actually roughly reach the sizes of actual devices. All three components of fields were taken into account. The code was developed in Matlab R2019a and run on Intel Xeon nodes with 190~GB RAM. Refractive indices of LC were set to $n_o = 1.5$, $n_e = 1.8$ \cite{li2004temperature} and the index of the surrounding isotropic material to $n_o = 1.5$, to match the ordinary refractive index of LC.

\bibliography{beam_steering}

\section*{Acknowledgements (not compulsory)}

This work was supported by the Slovenian Research Agency ARRS under Contracts No. P1-0099, No. J1-2462, and No. J1-1697. D.S. acknowledges support from the Ministry of Education, Science and Sport, EU Cohesion Policy, under Contract Raziskovalci-2.1-UL-FMF-952012 and from Adria Tehnik d.o.o..

\section*{Author contributions}
D.S. conceived and led the work. U.M. and D.S. performed numerical simulations, supervised by M.R. All authors analysed the results and contributed in writing the manuscript.

\section*{Competing interests}
Authors declare no potential conflict of interest.

\section*{Additional information}
Correspondence and requests for materials should be addressed to D.S.

\end{document}